\documentclass[aps, prl, reprint, a4paper,superscriptaddress,nobibnotes]{revtex4-1}
\usepackage{amsmath}
\usepackage{color}
\usepackage{graphicx}

\usepackage[hidelinks]{hyperref}

\usepackage[left=1.6cm,right=1.6cm,top=3cm,bottom=3cm]{geometry}
\begin{document}


\title{RF-dressed Rydberg atoms in hollow-core fibres}
\author{C. Veit}
\affiliation{5. Physikalisches Institut and Center for Integrated Quantum Science and Technology, Universit\"at Stuttgart, Pfaffenwaldring 57, 70550 Stuttgart, Germany}
\author{G. Epple}
\affiliation{5. Physikalisches Institut and Center for Integrated Quantum Science and Technology, Universit\"at Stuttgart, Pfaffenwaldring 57, 70550 Stuttgart, Germany}
\affiliation{Max Planck Institute for the Science of Light, G\"unther-Scharowsky-Str. 1/Bldg. 24, 91058 Erlangen, Germany}
\author{H. K\"ubler}
\affiliation{5. Physikalisches Institut and Center for Integrated Quantum Science and Technology, Universit\"at Stuttgart, Pfaffenwaldring 57, 70550 Stuttgart, Germany}
\author{T. G. Euser}
\affiliation{Max Planck Institute for the Science of Light, G\"unther-Scharowsky-Str. 1/Bldg. 24, 91058 Erlangen, Germany}
\affiliation{Cavendish Laboratory, JJ Thompson Ave, University of Cambridge, Cambridge, CB3 0HE, UK}
\author{P. St. J. Russell}
\affiliation{Max Planck Institute for the Science of Light, G\"unther-Scharowsky-Str. 1/Bldg. 24, 91058 Erlangen, Germany}
\author{R. L\"ow}
\email{r.loew@physik.uni-stuttgart.de}
\homepage{www.pi5.uni-stuttgart.de}
\affiliation{5. Physikalisches Institut and Center for Integrated Quantum Science and Technology, Universit\"at Stuttgart, Pfaffenwaldring 57, 70550 Stuttgart, Germany}


\begin{abstract}
The giant electro-optical response of Rydberg atoms manifests itself in the emergence of sidebands in the Rydberg excitation spectrum if the atom is exposed to a radio-frequency (RF) electric field. Here we report on the study of RF-dressed Rydberg atoms inside hollow-core photonic crystal fibres (HC-PCF), a system that enables the use of low modulation voltages and offers the prospect of miniaturised vapour-based electro-optical devices. Narrow spectroscopic features caused by the RF field are observed for modulation frequencies up to $500\, \text{MHz}$.
\end{abstract}


\maketitle 

\section{Introduction}
The large polarisability of Rydberg states~\cite{Gallagher05} enhances the electro-optical response of a thermal Rydberg gas. The combination of thermal vapours and the spectroscopically narrow features of electromagnetically induced transparency involving Rydberg states~\cite{MohapatraAdams07} allows electro-optical effects to occur at very small electric field amplitudes. The interplay between Rydberg atoms and radio-frequency (RF) or microwave electric fields is especially interesting, since its nature strongly depends on many different parameters such as the modulation frequency, the Kepler frequency, the excitation Rabi frequencies and potentially the mutual interaction strength between the Rydberg atoms themselves. All these parameters can be made comparable in their magnitude in the experiment and tuned over a wide range, permitting exploration of different manifestations of the field-atom interaction. Ac-dressed Rydberg atoms have been employed to study microwave excitation and ionization~\cite{BayfieldSharma81,HeuvellGallagher84}, to demonstrate phase modulation of light~\cite{MohapatraAdams08}, to measure atomic polarisabilities~\cite{ZhangLeventhal94,BaughLeventhal96} and to observe St\"uckelberg oscillations in cold atomic ensembles~\cite{DitzhuijzenLinden09}. Regarding potential future applications, it is especially interesting that RF-dressed Rydberg atoms have been shown to exhibit enhanced dc electric field sensitivity~\cite{BasonAdams10} and that Rydberg atoms also allow accurate sensing of microwave electric fields~\cite{SedlacekShaffer12,SedlacekShaffer13,AndersonRaithel16}, strong RF fields~\cite{JiaoJia16,MillerRaithel16} and RF noise~\cite{ZhelyazkovaHogan15,ZhelyazkovaHogan15b}.

Adopting the approach presented in~\cite{BasonAdams10}, we study RF-dressed Rydberg states and reduce the size of the initial macroscopic vapour cell experiment to the length-scales of a hollow-core photonic crystal fibre (HC-PCF)~\cite{Russel03,EppleLoew14}. In this way we can not only minimize the modulation voltage and the power consumption but also increase the bandwidth. The performance of the system is analysed by the position and magnitude of sidebands in the Rydberg excitation spectrum that appear when the RF field is applied. Throughout this article we discuss a regime in which the applied modulation frequencies and powers justify a perturbative Floquet analysis of the spectroscopic features~\cite{ZhangLeventhal94,DitzhuijzenLinden09,BasonAdams10}.

\section{Theoretical description}
The extreme sensitivity of highly excited atoms to external fields allows modification of the absorptive and dispersive behaviour of a thermal gas by weak applied electric fields. In the following we consider the effect of a RF field (with dc offset)
\begin{equation}
\label{eq:efield}
E(t) = E_\text{dc} + E_\text{ac} \, \text{cos}\,\omega t \, ,
\end{equation}
on a long-lived Rydberg state ($\Gamma_\text{ryd}<<\omega$). We restrict the analysis to a regime where the modulation frequency $\omega$ is larger than the Rabi frequencies of the excitation lasers but also much smaller than the transition frequencies to neighbouring Rydberg states. This implies that the atom-light interaction cannot follow the energy modulation of the Rydberg state, enabling us to treat the external field as quasi-static. It is therefore justified to apply time-independent perturbation theory at sufficiently low field amplitudes. In this regime, the Rydberg state is split into equidistant quasi-stationary Floquet states separated by multiples of the modulation frequency~\cite{BasonAdams10}. The considered $P$-state of caesium experiences the quadratic Stark effect causing the energy of the Rydberg state $\varepsilon(t)$ to shift according to:
\begin{equation}
\label{eq:energy1}
\varepsilon(t)    = \varepsilon_0 - \frac{1}{2} \alpha E(t)^2  \, .
\end{equation}
Here $\varepsilon_0$ is the energy and $\alpha$ the polarisability of the unperturbed state. Inserting the electric field from Eq.~\ref{eq:efield} we obtain the expression
\begin{equation}
\label{eq:energy2}
\varepsilon(t) = \varepsilon_\text{s} - \alpha \left[ E_\text{dc} E_\text{ac} \text{cos}\,\omega t  + \frac{E_\text{ac}^2}{4} \text{cos}\,2\omega t\right] \, ,
\end{equation}
consisting of a static contribution 
\begin{equation}
\label{eq:energystatic}
\varepsilon_\text{s} = \varepsilon_0 - \alpha (2 E_\text{dc}^2 + E_\text{ac}^2)/4 \, ,
\end{equation}
and terms oscillating at $\omega$ and $2\omega$. If the admixture of neighbouring Rydberg states is small, the electron wave function $\Psi(\mathbf{r},t)$  can be separated into a time-independent part $\psi(\mathbf{r})$ and a temporally varying part determined by Eq.~\ref{eq:energy2}. The time-dependent Schr\"odinger equation can then be integrated to obtain \cite{DitzhuijzenLinden09,ZhangLeventhal94}
\begin{equation}
\label{eq:wavefunction}
 \Psi(\mathbf{r},t) = \sum_{n=-\infty}^{\infty } A_\text{n}(x,y) \, \text{exp}\left( -\frac{i}{\hbar} \varepsilon_\text{s} t + in\omega t\right) \psi(\mathbf{r}) \, ,
\end{equation}
where the coefficients
\begin{equation}
\label{eq:an}
 A_\text{n}(x,y)=\sum_{m=-\infty}^\infty J_\text{n-2m}(x) J_\text{m}(y) 
\end{equation}
are determined by Bessel functions of the first kind and
\begin{equation}
\label{eq:xy}
x = \frac{\alpha E_\text{dc} E_\text{ac}}{\hbar \omega} \qquad \text{and} \qquad  y = \frac{\alpha E_\text{ac}^2}{8\hbar \omega}\, .
\end{equation}
\begin{figure}[t!]
   \includegraphics[scale=1]{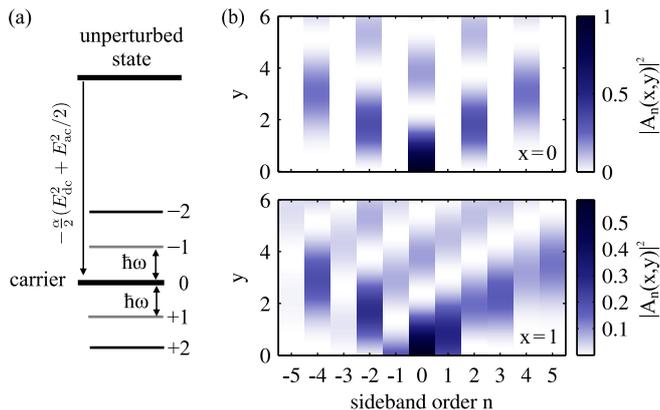}
    \caption{(a) In the presence of a RF electric field, the originally unperturbed Rydberg state is shifted and develops sidebands with relative amplitude $|A_\text{n}(x,y)|^2$ (see Eq. \ref{eq:an}) at multiples of the RF frequency. (b) For a vanishing dc field ($x=0$), only even-order sidebands are populated, whereas for $x\neq0$ also odd-order sidebands occur and the sideband spectrum becomes asymmetric.}
   \label{fig:fig1}
\end{figure} 
Equation \ref{eq:wavefunction} predicts not only a constant shift of the originally unperturbed Rydberg state but also the formation of sidebands with a fixed phase relationship and energies
\begin{equation}
\label{eq:sbenergies}
\varepsilon_\text{n}= \varepsilon_\text{s} - n \hbar \omega = \varepsilon_0 - \alpha (2 E_\text{dc}^2 + E_\text{ac}^2)/4 - n \hbar \omega\, .
\end{equation}
If optically probed, these sidebands, illustrated in Fig.~\ref{fig:fig1}(a), appear in the absorption spectrum and can be interpreted as being caused by multi-photon excitation involving optical excitation photons and $|n|$ RF photons. Correspondingly, stimulated emission involving various numbers of RF photons can lead to modulation of the excitation light, which as a result acquires sidebands. It is worth noting that even for a pure ac field ($E_\text{dc}=0$), the root-mean-square amplitude of the ac field causes a shift of the carrier state ($n=0$) according to $\varepsilon_s = \varepsilon_0 -\frac{\alpha}{4} E_\text{ac}^2$.

From Eq. \ref{eq:wavefunction} we can infer that the relative amplitude of a specific sideband is given by the expression $|A_\text{n}(x,y)|^2$, which vanishes for sufficiently large $n$. $|A_\text{n}(x,y)|^2$ is illustrated in Fig.~\ref{fig:fig1}(b) as a function of $y$ for both finite and vanishing dc fields. For even-order sidebands and $E_\text{dc}=0$, the sum in Eq.~\ref{eq:an} reduces to a single term, since $J_\nu(0)$ is nonzero only for $\nu=0$. For odd-order sidebands, the sum vanishes completely for $E_\text{dc}=0$. This is consistent with Eq.~\ref{eq:energy2}, which predicts modulation of the energy at $\omega$ only in the presence of a dc field. 

For $E_\text{dc}=0$, it is particularly easy to determine the ac field amplitude that maximizes the amplitude of a specific sideband at a given driving frequency. For the second-order sidebands, for example, the relative sideband amplitude is given by $|A_2(0,y)|^2 = |J_1(y)|^2$ and reaches a maximum of $\sim 33.9\%$ at $y\approx 1.84$. For the $30P_{3/2}$ ($m_\text{J}=1/2$) state (considered later) and a modulation frequency of $500\, \text{MHz}$, this corresponds to an ac field amplitude $E_\text{ac}\approx 16.5\, \text{V/cm}$. Since $y$ is at the same time proportional to the inverse of the modulation frequency $\omega$ and $E_\text{ac}^2$ (Eq.~\ref{eq:xy}), the RF voltage required to achieve efficient modulation scales with the square root of the modulation frequency.
\section{Experimental setup and excitation scheme}  
\begin{figure}[t!]
    \includegraphics[scale=1]{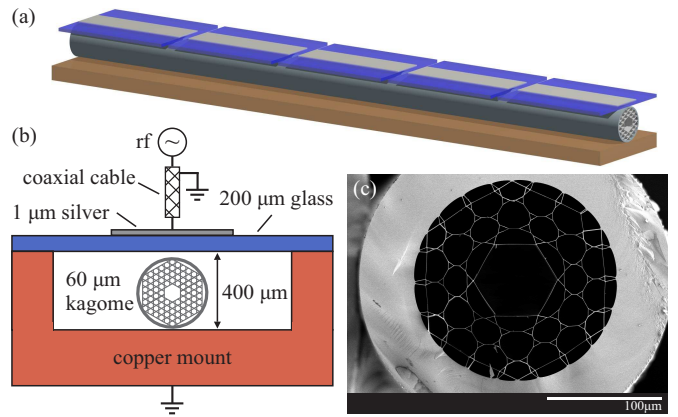}
    \caption{(a) A hollow-core PCF with a core diameter of $60\,\mu\textrm{m}$ (not to scale) is mounted inside a vacuum chamber exposed to caesium vapour. Five field plates allow individual sections along the fibre to be addressed with ac electric fields. In the experiments, only one field plate is modulated at a time. (b) The field plates consist of $2\,\textrm{mm}$ wide silver stripes evaporated on to thin glass substrates and then positioned above the fibre. (c) Microstructure of the kagom\'e-PCF with a core diameter of $60\,\mu\textrm{m}$.}
    \label{fig:fig2}
\end{figure}  
\begin{figure*}[t!]
\includegraphics[scale=1]{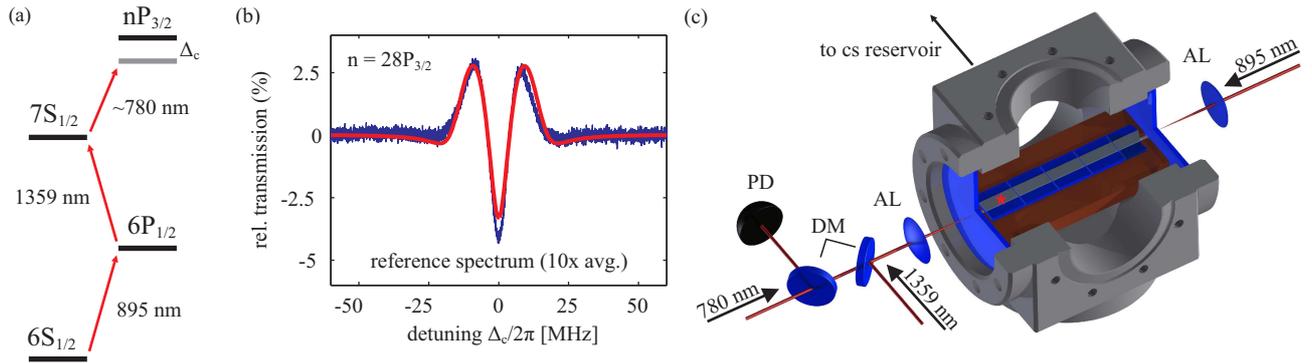}
    \caption{(a) Three-photon transition to the Rydberg state. During measurements, both the $895\,\textrm{nm}$ and the $1359\,\textrm{nm}$ laser were resonantly locked and the $780\,\textrm{nm}$ Rydberg laser was scanned. The transmission of the $895\,\textrm{nm}$ probe laser was measured as a function of the Rydberg laser detuning $\Delta_\text{c}$. (b) Typical probe transmission curve obtained in a reference cell (blue). The experimental data can be accurately modelled using a density matrix approach that accounts for the magnetic sublevels and the thermal velocity of the atoms (red curve). (c) The excitation beams are combined using dichroic mirrors (DM) and focused into the fibre ends with achromatic lenses (AL). The transmission of the $895\,\textrm{nm}$ probe beam is detected with a photodiode (PD). For the measurements, described here, only one field plate (red mark) was modulated.}
    \label{fig:fig3}
\end{figure*} 
We study RF-dressed Rydberg atoms inside a vapour-filled kagom\'e-type HC-PCF (core diameter $60\,\mu\textrm{m}$ - see Fig.~\ref{fig:fig2}(c)). By confining both excitation light and atoms, these broadband-guiding fibres offer an elegant way to interface atomic vapours with light. The fibre is mounted inside a CF63 vacuum cube connected to a flexible metal bellow acting as caesium reservoir. After pumping and baking out the vacuum system, the valve connecting the chamber to the turbo pump was closed and a caesium ampoule inside the reservoir bellow was broken. The caesium then diffused into the main chamber and on into the open ends of the fibre. By changing the reservoir temperature, we are able to control the vapour density inside the main chamber. As explained later in more detail, due to slow diffusion of the atoms the vapour densities inside and outside the fibre are not at equilibrium.

As illustrated in Fig.~\ref{fig:fig2}(a), the fibre is equipped with five field plates, allowing individual sections along the fibre to be addressed with electric fields. The electrodes were evaporated on to $200\,\mu\text{m}$ thick glass substrates as $2\,\text{mm} \times 22\,\text{mm}$ silver stripes with a thickness of $\sim 1\, \mu\text{m}$ (see Fig.~\ref{fig:fig2}(b)). A $5\,\textrm{nm}$ thick layer of chromium, deposited between the glass and the silver layer, ensured a strong bond between the glass and the electrodes. For reasons of compatibility with the reactive alkali gas, the field plates were electrically connected using in-house-fabricated coaxial transmission lines consisting of a central conductor surrounded by alumina beads and a nickel-plated copper braiding. Despite the impedance mismatch between the in-vacuum cables ($\sim28\,\Omega$) and the air-side cables ($50\,\Omega$), this setup allows RF frequencies of over $2\, \text{GHz}$ to be applied while requiring only low modulation voltages due to the miniaturised design.

We excite the Rydberg state via the three-photon transition $|6S_{1/2},\,F=3\rangle \longrightarrow |6P_{1/2},\,F'=4\rangle \longrightarrow |7S_{1/2},\,F''=4\rangle \longrightarrow |nP_{3/2}\rangle$, respectively featuring the transition frequencies $895\,\textrm{nm}$, $1359\,\textrm{nm}$ and $\sim780\, \textrm{nm}$ (see Fig. \ref{fig:fig3}(a)). As shown in Fig.~\ref{fig:fig3}(c), the excitation beams are combined via dichroic mirrors and focused into the fundamental mode of the fibre with achromatic lenses. The mode field diameter inside the fibre is approximately $40\,\mu\text{m}$ and the in-out efficiencies of the three beams are $\eta_{895}\approx 54\%$, $\eta_{780}\approx 55\%$ and $\eta_{1359}\approx 33\%$, where the subscripts indicate the wavelength. The polarization of all beams is parallel to the applied RF field. In order to minimize Doppler broadening, the $895\,\textrm{nm}$ probe beam is arranged to counter-propagate relative to the two other beams. During measurements, the two lasers driving the lower transitions were frequency-stabilized by means of Doppler-free DAVLL spectroscopy \cite{petelski2003} and two-photon polarisation spectroscopy \cite{carr2012polarization}, while the $780\,\textrm{nm}$ laser was scanned across the Rydberg line. A photodiode detected the transmitted probe light, which we measured as a function of the Rydberg laser detuning $\Delta_\text{c}$. We scaled the detuning axis using a Fabry-P\'erot cavity that was actively length-stabilised via the locked probe laser.

Figure~\ref{fig:fig3}(b) shows the typical transmission spectrum of the probe field measured in a reference cell. Qualitatively, the two transmission peaks in the signal result from the two dressed states emerging when the intermediate transition is diagonalised. For a more quantitative understanding of the lineshape, the thermal velocity of the atoms must be taken into account \cite{carr2012}. We  modelled the atomic system with a density matrix approach using the Liouville-von Neumann equation. In the simulations we included the hyperfine magnetic sublevels $m_\text{F}$ of the lower three states, a single Rydberg state for each magnetic quantum number (the hyperfine splitting of the Rydberg state is not resolved in the experiment), the dark hyperfine ground state ($6S_{1/2},\,F=4$) and transit time effects via an additional decay channel for all states. We calculated the density matrix for all relevant velocity classes of the atoms and obtained a Doppler-averaged density matrix $\rho$ by weighting the results with a Boltzmann distribution. Within the applied framework, the absorption of a two-level system is determined by the imaginary part of the coherence between the two states. For the multi-level system considered here, the absorption of the probe laser is determined by $\text{Im}\rho_{21}$, representing the sum of all coherences between the magnetic substates of the ground state and the magnetic substates of the first excited state. In this sum, every coherence is weighted with the corresponding transition dipole matrix element. We related our simulations to the experimental data by fitting the expression $t = a \cdot \text{exp}(b\, \text{Im}\rho_{21})$ to the transmission spectra (treating $a$ and $b$ as free fitting parameters) and obtained good agreement (see Fig.~\ref{fig:fig3}(b)).
\section{Results} 
\begin{figure}[t!]
    \includegraphics[scale=1]{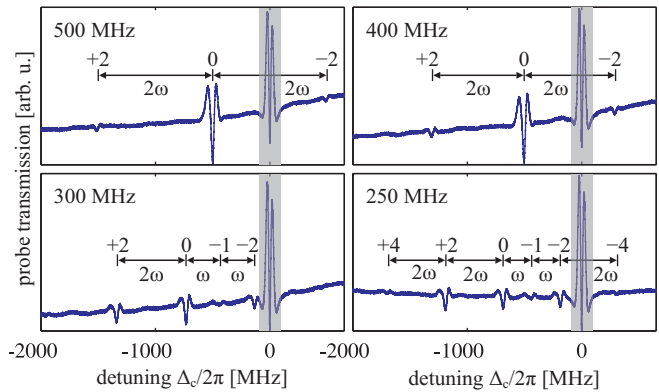}
    \caption{Measured transmission of the probe beam at different RF modulation frequencies. In all traces, the largest signal (shaded grey) originates from the unmodulated parts of the fibre. The field amplitudes were estimated from the red-shift of the carrier state ($n=0$) and vary slightly from each other due to the transmission properties of the in-vacuum coaxial cable. RF-induced sidebands are clearly visible on both sides of the carrier state. At high modulation frequencies only the $\pm2$ orders are visible, whereas at lower modulation frequencies more sidebands appear. The existence of odd-order sidebands is attributed to local electric fields inside the fibre, which are estimated to be smaller than 5\% of the applied ac field.}
    \label{fig:fig4}
\end{figure}
In the following we present spectroscopic results obtained in the hollow-core fibre while a single field plate (marked in Fig.~\ref{fig:fig3}(c)) was modulated at frequencies up to $500\,\text{MHz}$. The fibre had been exposed for 5 weeks to a Cs atmosphere at different chamber temperatures of up to $90\,^{\circ}\text{C}$. The reservoir temperature was always kept below the chamber temperature. The optical density (of the resonant probe transition) measured inside the fibre was $O\!D_\text{f}=4.4$ and smaller than the optical density of the surrounding chamber $O\!D_\text{ch} = 19$. Due to the mismatch in optical density, which as mentioned originates from the slow diffusion of the atoms into the fibre, we do not expect the atomic number density to be uniform along the fibre at the time of the measurements. In fact, a Stark-shifted signal was only observed when either one of the two outermost plates was modulated. Future studies may reveal whether the lack of signal from the inner sections of the fibre is due to the absence of Cs atoms or to high electric fields prohibiting Rydberg excitation. In all the measurements, the chamber and reservoir temperatures were $T_\text{ch}\approx 84\,^{\circ}\text{C}$ and $T_\text{res}\approx 47\,^{\circ}\text{C}$ respectively. The modulation potential was purely sinusoidal (i.e. no dc offset was applied) and the externally supplied RF voltage was $U_\text{rf}=250\,\text{mV}$ ($\sim 1\,\textrm{dBm}$). At the input face of the fibre, the beam powers were $p_{895}= 1\,\mu\textrm{W}$, $p_{1359}=8.2\,\mu\textrm{W}$ and $p_{780}=25.5\,\textrm{mW}$. The excited Rydberg state $30P_{3/2}$ ($m_\text{J} = \pm1/2$) possesses a polarisability \mbox{$\alpha_{30P_{3/2}} = 26.93\,\textrm{MHz/(Vcm$^{-1}$})^2$ \cite{gu1997}.}

Figure~\ref{fig:fig4} shows the probe transmission (averaged over 500 traces) for four different modulation frequencies between $250\,\textrm{MHz}$ and $500\,\textrm{MHz}$ as a function of $\Delta_\text{c}$. In all the traces, the largest signal (marked by shaded area) originates from the unmodulated part of the fibre, through which the probe beam passed before it encountered the modulated section at the far end of the fibre (see Fig.~\ref{fig:fig3}(c)). Due to the high probe power employed, the signal is broadened and the depth of the central absorption dip is smaller than in the weak-probe limit. In the modulated section of the fibre, the atomic resonance (corresponding to $n=0$ in Eq.~\ref{eq:sbenergies}) is red-shifted. The shifted Rydberg state, which we refer to as the carrier, gives rise to a distinct and clear signal in the spectrum at negative detunings $\Delta_\text{c}$. Setting $E_\text{dc}=0$ we used Eq.~\ref{eq:energystatic} to estimate the field amplitude $E_\text{ac}$ from the red-shift of the carrier state. Due to the frequency response of the coaxial transmission line we obtained slightly different field amplitudes for the different modulation frequencies lying between $\sim 8.6\,\textrm{V/cm}$ at $500\,\textrm{MHz}$ and $\sim10.1\,\textrm{V/cm}$ at $250\,\textrm{MHz}$. From the geometry of the setup we conclude from these results that, at a modulation frequency of $500\,\textrm{MHz}$, approximately 70\% of the input RF voltage is transferred to the field plates. This is consistent with the impedance mismatch between the air-side coaxial cable and the self-built in-vacuum transmission line.

As expected, the carrier state is accompanied by RF-induced sidebands separated by multiples of the RF frequency. At high modulation frequencies, only second order sidebands can be identified and the carrier state possesses the largest amplitude. At lower modulation frequencies, higher-order sidebands also appear and the carrier amplitude falls. For the second order sidebands, relative amplitudes of $\sim 5\%$ and $\sim29\%$ are reached at modulation frequencies of $500\,\textrm{MHz}$ and $250\,\textrm{MHz}$ respectively. Strikingly, the sideband spectrum becomes asymmetric in the case of low modulation frequencies and also odd-order sidebands occur. Since no dc field was applied, both these effects must be attributed to local electric fields inside the fibre. This observation confirms previous findings~\cite{EppleLoew14}, where line-shifts indicated the presence of local electric fields inside a kagom\'e-PCF with a core diameter of $60\,\mu\textrm{m}$. These shifts seemed to vanish after long-term exposure of the fibre to a Cs atmosphere~\cite{EppleLoew14}. 
\begin{figure}[t!]
    \includegraphics[scale=1]{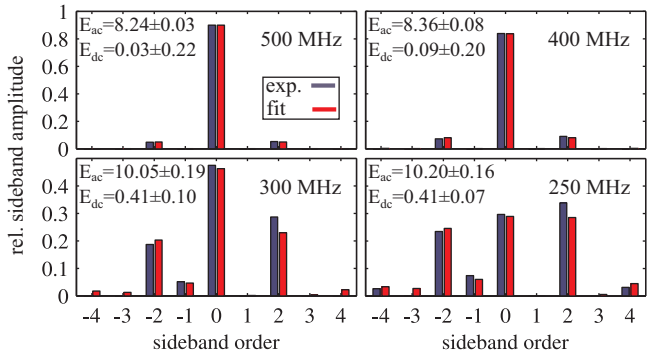}
    \caption{Relative sideband amplitudes for the data-set shown in Fig.~\ref{fig:fig4}. The blue bars represent experimental data determined from the spectra. The red bars represent a fit of the theoretically expected amplitude $|A_\text{n}(x,y)|^2$. The field amplitudes (given in V/cm) are extracted from the fit parameters $x$ and $y$ (see Eq.~\ref{eq:xy}) and agree within 4\% with the ac field amplitudes estimated from the red-shift of the carrier state. Despite its low amplitude ($<0.5\,\textrm{V/cm}$), the local dc field inside the fibre induces considerable asymmetry.}
    \label{fig:fig5}
\end{figure} 

To compare the experimental results with theoretical predictions and to estimate the local dc field inside the fibre, we fitted $|A_\text{n}(x,y)|^2$ (see Eq.~\ref{eq:an}) to the experimentally obtained sideband amplitudes of the data-set shown in Fig.~\ref{fig:fig4}. We then extracted the ac and dc field from the fit parameters $x$ and $y$ (see Eq.~\ref{eq:xy}). This procedure assumes that the ac and dc fields are parallel, while the dc field inside the fibre is likely to be randomly oriented. Since, however, the tensor polarisability of the $30P_{3/2}$ state is smaller than 10\% of the scalar polarisability \cite{gu1997}, we are confident that a good estimate of the field strengths can be obtained. Figure~\ref{fig:fig5} shows both the experimental data and the fits, which are in excellent agreement with each other. The fitted amplitudes $E_\text{ac}$ correspond within 4\% to the values estimated from the red-shift of the carrier state. For the dc fields, we obtained field strengths smaller than $0.5\,\textrm{V/cm}$, which is again consistent with previous results~\cite{EppleLoew14}. In the parameter regime investigated, the sideband spectrum becomes relatively insensitive to small dc fields at large modulation frequencies. As a result, the estimated field amplitudes $E_\text{dc}$ have relatively large uncertainties in the high frequency case (see Fig.~\ref{fig:fig5}). Nevertheless, it is striking that the local dc field experienced by the atoms seems to increase with decreasing modulation frequency. While this behaviour is not yet understood, further investigations are expected to yield valuable insight into the charge dynamics inside the fibre.
\section{Outlook}
In future experiments we plan to exploit the spatial resolution of patterned electrodes to obtain a more detailed understanding of diffusion, adsorption and desorption effects inside the fibre core. Also promising is the integration of the modulation electrodes in the form of conductive microwires inside the fibre~\cite{TyagiRussel10}. This will allow even smaller modulation voltages and will lead to larger bandwidths. Due to the tensor polarisability of $P_{3/2}$ states, the magnitude of the individual sidebands can provide a more detailed insight into the direction of the electric fields and the distribution of ions potentially sitting on the walls of the hollow fibre core. 

So far we have observed and analysed the applied radio-frequency fields only in terms of additional field-dependent spectroscopic features. If an efficient, miniaturised electro-optical device is to be realised, the back-action of the atomic system on the excitation light must be investigated. This can be carried out by measuring the amplitude of the optical sidebands (e.g., using a heterodyne setup). The kagom\'e-PCF system also allows the study of the atomic response in a regime where a perturbative approach is insufficient. This could be achieved either by applying microwaves resonant to Rydberg-Rydberg transitions or by increasing the power to a level where the Rabi frequencies become comparable to the transition frequencies between Rydberg states. In both cases the optical response of the system could then be studied in situations where the Rydberg electron cannot adiabatically follow the applied fields. 
\begin{acknowledgments}
We thank G. Untereiner for the vapour deposition of the field plates and G. Raithel for discussions. This project was
financed by the Baden-W\"urttemberg Stiftung and supported by the ERC under contract number 267100.  H. K. acknowledges funding from the Carl-Zeiss Foundation. G. E. acknowledges funding from the Erlangen-based International Max Planck Research School: Physics of Light (IMPRS-PL). C. V. was supported by the Deutschlandstipendium.
\end{acknowledgments}
\bibliography{RfDressing}
\end{document}